# Overview of Surgical Simulation

Mohamed A. ElHelw

Center for Informatics Science
Nile University
Giza, Egypt

melhelw@nu.edu.eg

Motivated by the current demand of clinical governance, surgical simulation is now a well-established modality for basic skills training and assessment. The practical deployment of the technique is a multi-disciplinary venture encompassing areas in engineering, medicine and psychology. This paper provides an overview of the key topics involved in surgical simulation and associated technical challenges. The paper discusses the clinical motivation for surgical simulation, the use of virtual environments for surgical training, model acquisition and simplification, deformable models, collision detection, tissue property measurement, haptic rendering and image synthesis. Additional topics include surgical skill training and assessment metrics as well as challenges facing the incorporation of surgical simulation into medical education curricula.

## 1. Introduction

Surgical training is an ongoing process where the constant evolution of surgical techniques makes it no longer possible to master a single set of skills as a trainee surgeon and rely on them for the entire career. Advances in surgical techniques are inseparably linked to advances in surgical technology and the pace of change is constantly accelerating. Moreover, most aspects of medicine have historically been learnt in an apprenticeship model by means of observation, imitation, and instruction. In such a setting, much of the expertise transferred from the mentor to the trainee is implicit and cannot be transferred easily to a didactic setting. Thus far, there are few standardized training methods in surgery. Commonly regarded essential competencies include manual dexterity, familiarity with high-tech equipment, sound professional judgement and the ability to integrate technical skills with clinical practices. In general, training solutions need to be developed in tandem with the surgical techniques they aim to teach.

Virtual Reality (VR), or Virtual Environments (VE), -based simulators are rapidly becoming an integral part of surgical training and skills assessment. Current high-fidelity simulators offer the opportunity for safe, repeated practice and objective measurement of performance. However, for deploying the technology or the development of future simulators, it is important to look at how the history and clinical requirements of surgical training have evolved in the past. In addition, several technical issues related to the creation of surgical VR-based simulators have to be considered including model acquisition and simplification, deformable models [1], collision detection [2], tissue property acquisition [3] and haptic rendering. A significant amount of research has also been carried out in the creation of realistic rendering, *i.e.* realistic image synthesis, especially those focusing on non-rigid organs [4] and using hardware accelerated rendering [5]. Additional topics to consider include types of skill training and assessment metrics as well as challenges facing the incorporation of surgical simulation into medical education curriculum. In this paper we provide an overview of the key topics involved in surgical simulation and associated technical challenges. It is worth noting that the paper is a part of an old thesis and most of the references are thus not up to date. However, the key areas that need to be considered in surgical simulation have not changed since then.

## 2. Clinical Motivation

Medical education is conventionally accentuating a curriculum based on cognitive, psychomotor, and affective domains of learning which were originally proposed nearly 50 years ago [67]. However, the incongruity between evidence-based recommendations and real-world practice highlights the inadequacy of the preceptored medical education tradition [68]. Consequently, there has been a shift in the method of medical education towards experiential ('hands-on') medical learning.

### 2.1. Operating Theatre Apprenticeship

Conventional surgical training is based on the apprenticeship model. Under the scrutiny of experienced instructors, surgical trainees learn by observing, then gradually performing specific procedures inside the operating theatre. The theoretical knowledge of the process is assumed to be gained beforehand through learning. Although the operating theatre is a basic element of surgical training, it is becoming less effective due to several factors. First, trainees are exposed to heterogeneous distribution of procedures depending on the flow of patients; it is also time consuming and costly. This can result in large variations in the professional standards of surgeons. Moreover, operating theatre-based training can constitute a potential risk to patients due to the inevitable distraction while training on complex or advanced procedures. For these reasons, several out-of-the-operating-theatre training approaches have been considered.

### 2.2. Computer-Based Training

With the current advances in computer hardware and Internet technology, the level of computer literacy and the dependency on computerized information are steadily increasing. One of the established training methods is based on interactive multimedia applications where the trainee interacts visually with the system in order to learn the necessary steps involved in certain surgical procedures. Existing research has shown that Computer-Based Training (CBT) improves the teaching efficiency and can significantly cut down on learning times. Nevertheless, it is still inadequate for providing effective training on basic surgical skills. This is largely due to the difficulty of imitating surgical procedures by using the Two-Dimensional (2D) CBT systems that have limited immersion, physical control and interaction.

### 2.3. Animals and Cadavers

Students traditionally learn the basics of surgery on live lab animals. For example, the use of surgical instruments begins with anatomical dissections and physiological experiments. Anesthetised animals, typically dogs, pigs or rats, have long been a part of the curriculum in medical schools [6]. However, this training modality is becoming increasingly unacceptable due to obvious ethical, legal and humane concerns. Life animals shouldn't be sacrificed for teaching purposes, and some infectious diseases can be transmitted from animals to trainees. From a technical point of view, the anatomy of animals and human beings are different, and the cost associated with preserving and disposing of used models is high.

### 2.4. Synthetic Models

Synthetic models provide a low-risk training opportunity. Many models resembling parts of the human anatomy with complexity ranging from bone structures to fully integrated models are available. In Figure 1 some examples of synthetic models are illustrated, showing detailed head, knee arthroscopy, and central venous catheterisation models that are currently available. These models are typically made form plastic or latex materials and can be used to teach basic examination methods, tissue dissections, and suturing. In practice, training with synthetic models is usually used during preliminary stages of surgical education. This modality, however, has limited realism and it is also difficult and costly to acquire and maintain a large number of different cases [7].

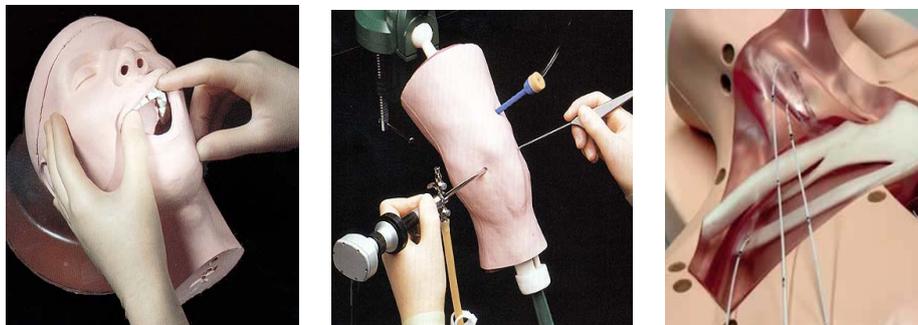

**Figure 1.** Examples of synthetic models used for surgical training include (a) a detailed head model for practising trans-oral and dental surgeries, (b) a knee arthroscopy and open surgery trainer [8], and (c) a model for teaching central venous catheterization [9].

## 3. Surgical Training and Virtual Environments

Over the last twenty-five years, there has been a strong movement towards changing the traditional approaches to surgical training. One of the major drivers for the development of surgical simulation is the advent of new surgical practices such as interventional radiology and MIS techniques. Standard MIS procedures, for example bronchoscopy and laparoscopy, are carried out by inserting instruments through natural body openings or small artificial incisions and the operating field is observed on a 2D screen. If handled properly, MIS instruments are harmless, with patient trauma and hospitalization greatly reduced and diagnostic accuracy and therapeutic success enhanced. The introduction of MIS, however, has placed new requirements to conventional surgical training. The complexity of instrument controls, restricted vision and mobility, difficult hand-eye coordination, unnatural perceptual-motor relationships, and the lack of tactile perception require a high degree of manual dexterity and coordination from the operator. Consequently, much attention has been directed to new training methods such as surgical simulators for acquiring these skills. However, the success of these simulators in early years was limited by the technology, lack of patient–specific information, and unrealistic tissue deformation.

Recently, advances in computer graphics have enabled the introduction of high quality VE for surgical training. Such systems are increasingly being used for practicing new or complex procedures, consolidating existing skills, and objective evaluation of certain measures of surgical competence. Before the development of surgical simulation systems, a number of issues have to be considered. For example, an important decision is whether the surgical process to be modelled is amenable for computer-based simulation. This in turn depends on the complexity of the process, accessible technology, as well as financial constraints. In general, issues that influence the selection of a process include [10]:

- Process complexity and practice frequency.
- Hazard to patients and trainees.
- Simulator effectiveness in terms of realism and skills transfer.

In order for surgical simulation to achieve its goals, it has to mimic the genuine process to a great extent. The unique nature of the modus operandi of MIS makes computer simulation an ideal candidate for its training. MIS procedures are characterized by a narrow Field Of View (FOV), constrained movement, modest force feedback, and 2D video display. As a result, they are more acquiescent to computer simulation based on existing technology. Simulators for needle-based procedures, catheters guide wires, and other small-bore devices have also been developed. These procedures have similar characteristics as those found in MIS, *i.e.* limited haptic feedback and reliance on 2D video for visual feedback. Open surgery procedures, on the other hand, are more difficult to simulate due to the complexity of hand/instrument-tissue interaction, large scale deformation, and the richness of sensory feedback [11]. In the next sections, we will discuss the acquisition of virtual models and the key elements of simulation environments.

## 4. Model Acquisition and Simplification

Detailed anatomical models are necessary to realistically simulate the visual appearance, motion constraints, and deformation behaviour of the anatomy. Such anatomical information is typically acquired from several medical imaging modalities that use electromagnetic waves, sound waves or magnetic fields. Examples from the first category include X-ray and Computed Tomography (CT) techniques where 2D semi-transparent projection views are generated for rigid skeletal anatomy. A sequence of CT images can be used to describe 3D structures to sub-millimetre detail. However, these methods have limited capability of soft tissue discrimination and can expose patients to harmful ionizing radiation. Ultrasnography techniques, on the other hand, use harmless ultrasound waves to generate echoes at organ boundaries and within tissues. Returned echoes to the transducer are detected, allowing cross-sectional images to be interactively displayed [12]. Recently, 3D ultrasound has also been introduced and its use in clinical practice is becoming widely available. The drawbacks of this imaging modality are due to the inherent limitation of the technique in tissue penetration and limited access window [13]. The introduction of Magnetic Resonance Imaging (MRI) constituted a significant advance in medical imaging. MRI has unrivalled capability in differentiating tissue types and can provide detailed anatomical, as well as functional information of the soft tissue. It is safe and can provide 2D cross sectional or 3D volumetric images in any orientation of the patient.

In order to generate a detailed geometric model from the imaging datasets, a segmentation step is required to identify and separate different parts of the anatomy of particular interest [12]. Existing segmentation algorithms can be divided into manual, automatic and semi-automatic approaches [14]. In manual segmentation, the anatomy to be segmented is manually marked in successive 2D slices or 3D volume datasets. This approach is accurate and insensitive to image noise or missing information. In practice, however, it is tedious and time consuming [15]. Automatic segmentation is fast and doesn't require operator intervention, but it suffers from accuracy setbacks dictated by the algorithms used. Semi-automatic segmentation combines the advantages of both approaches by using minimal human intervention to guide the segmentation process.

To obtain a geometric surface representation from the segmented volumetric data, a mesh extraction technique such as the marching cubes algorithm [16] can be employed. The result of this process is typically a dense polygonal mesh. Therefore, mesh reduction or simplification is required to enable interactive model manipulation. In general, polygonal simplification provides several advantages [17]. For example, it reduces storage, memory and transmission requirements. It can also be used to accelerate computation in procedures that require shape information such as those involved in deformation modelling, collision detection, and interactive scene rendering. In general, existing polygon simplification algorithms can be classified into different categories according to the following criteria [18]:

- *Topology*: topology-preserving algorithms maintain mesh connectivity and genus, thus resulting in high fidelity output. They are bounded by the need to maintain holes and require as input a mesh with manifold topology. Topology-modifying algorithms don't preserve mesh topology and thus can achieve significant simplifications with lower visual fidelity. As the marching cubes algorithm generates manifold meshes, topology-preserving simplification can be effectively used for mesh reduction.

- *Mechanism*: most simplification algorithms typically consider four polygon removal mechanisms. These mechanisms are sampling, adaptive subdivision, decimation and vertex merging.

- *Static, dynamic and view-dependent simplification*: mesh simplification can be carried out in pre-processing or interactively. In static methods, the simplification is carried out in pre-processing with the advantage of separating simplification from rendering. Dynamic methods, on the other hand, perform simplification on the fly and can use view dependency to decide upon the Level-of-Detail (LOD) to use for the current view.

## 5. Key Elements of Surgical Simulation

### 5.1. Deformable Models

In order to provide a realistic learning experience, the simulation environment has to faithfully reflect the characteristics of deformable tissues in terms of shape and behaviour. However, the interactive simulation of deformable surfaces represents a major challenge in developing surgical training platforms [1]. A representation of the surface is typically obtained as geometric mesh by the methods discussed earlier. The mesh elements are then interpreted as the physical elements, and hence accurate geometric models enable a proper physical behaviour. Thus far, different approaches to modelling deformations have been introduced and they can be divided into non-physical and physical-based methods. Whilst non-physical-based techniques use geometrical manipulations to rapidly deform an object, physical-based methods incorporate measured material properties of the tissue in the deformation process, thus allowing more accurate results.

### 5.2. Collision Detection

The problem of collision detection or interference determination has been the focus of research in a number of computer graphics applications including games, cloth simulation, animation, computer-aided design and interactive environments. It is a vital component of interactive virtual environments [2]. In surgical simulation, the aim of collision detection is to determine the spatial contact between virtual objects within the simulation environment. In general, collision detection is a computationally intensive task that requires checking for contact all the primitives of every object in the scene. For n primitives, the problem has a complexity of $O(n^2)$, thus it is considered as a major bottleneck in interactive simulations [19]. The factors that influence the selection of the collision detection algorithm for a particular application can be summarized as model representation, collision query types, and application specifications [20].

For conventional polygon-based representations, convex polytype-polytype (e.g. polyhedral models) intersections can be determined by computing the closest distance between pairs of points in the two models. For bounded rigid motion, the spatial and temporal coherence is exploited to reduce the number of collision inspections. It assumes that the relationship between objects in the simulation change insignificantly between consecutive time steps. Alternative techniques can be employed in case of general polytypes. For instance, contact can be detected by checking the intersection of convex bounding volumes that enclose the model. Simple volumes such as boxes and spheres are chosen for their fast intersection tests. Other types of bounding volumes have also been used for their desirable features. For example, Axis Aligned Bounding Boxes (AABB) [21] and Oriented Bounding Boxes (OBB) [22] are used for efficient update of volume coordinates and tighter fitting of volume extents, respectively. A hierarchy of bounding volumes such as AABB trees and OBB trees can be used to further improve the speed and accuracy of collision detection. In this case, the leaves of the hierarchy usually contain a single primitive such as a triangle, NURBS patch, polyhedron or any other entity that make up the object. The collision detection between a pair of objects is then divided into two phases: the broad phase (also known as the rejection test) where the intersection test is carried out for the bounding volumes, and the narrow phase (or exhaustive test) where the exact intersection of the primitives is performed to confirm collision and find the loci of intersection. Contact determination becomes more involved in case of non-polygonal model representations.

A number of image-based collision detection techniques have also been proposed where the graphics hardware is used to accelerate the intersection calculations [23, 24]. In this case, bounding volume techniques are used to find the overlapping region between the two objects to be checked for intersection. A depth image is then generated for that region and the interference test is carried along One-Dimensional (1D) intervals, where for each rendered object, the interval from the smaller to the larger depth values is assumed to approximately represent the object. An analogous intersection approach has been proposed for surgical simulation. In this case, a viewing frustum corresponding to the bounding volume of a surgical instrument is rendered and polygons contributing to rasterised pixels indicate tissue-instrument contact.

In addition to model representation, other factors determine the characteristics of a collision detection algorithm. For instance, different applications require different types of collision information in addition to the very basic query of whether the models are in contact. Colliding model parts, depth of penetration, the distance and time to collision in case of moving objects are examples of required information. Moreover, the nature of the simulation environment such as its dynamics, the number of objects to be checked for intersection, and whether they are rigid or deformable, has a marked impact on collision detection algorithms. In surgical simulation, object deformability and interactivity requirements make contact determination one of the major technical challenges. In this environment, contacts between deformable organs, surgical instruments, and self-collisions have to be efficiently computed. The issues to be considered in this case include [25]:

- Collisions and self-collisions: for realistic interaction between deformable objects, all contact points including those due to self-collisions have to be computed. Thus, collision detection algorithms that neglect self-collisions or assume rigid body nature cannot be used.

- Pre-processing: in accelerated collision detection algorithms that are based on hierarchical bounding volumes, the spatial data structures storing object bounds are constructed in pre-processing. As the topologies of deformable objects change during simulation, these structures have to be efficiently updated or even recreated.

- Collision information: precise collision information such as penetration depth is required for realistic collision response. In addition, the collision techniques have to provide continuous collision information; that is to detect the exact contact within successive time steps.

- Performance: interactivity is a key feature for real-time applications such as surgical simulation. In fact, the perceived quality of 3D interactive applications depends more on real-time collision response rather than the exactness of the simulation. Hence, for these applications, collision algorithms have to be fast with constant execution times.

### 5.3. Tissue Property Acquisition

In MIS procedures, the tissue involved is highly deformable and the simulator has to accurately replicate its behaviour. Therefore, the knowledge of biomechanical tissue properties is essential. This is also important for intra-operative planning and surgical guidance. However, computing tissue properties is a difficult problem as they vary from subject to subject and have a high degree of intra-subject inhomogeneity. Several techniques for acquiring tissue properties by measuring force/displacement or stress/strain responses have been proposed. Given the observed response and known boundary conditions, deformation parameters can be extracted by assuming some models of deformation and performing inverse calculations or minimization to determine these parameters. Nevertheless, significant differences are expected in the mechanical properties of in vitro and in vivo specimens. Ideally, measurements should be carried out in vivo but due to operational constraints imposed, most measurements are carried out in vitro.

### 5.3.1. In Vitro Methods

Thus far, most of the available tissue property information was obtained in vitro, i.e., tissues are modelled by determining their mechanical properties experimentally. For this process, different forces, with ranges similar to that of real procedures, are exerted on controlled tissue shapes with well-defined boundary regions and the resulting displacements recorded. Material parameters are then computed from the plotted force/displacement or stress/strain, relationships. Analogous methods can be used to determine tissue reaction to forces exerted by surgical instruments during certain procedures such as cutting or grasping.

### 5.3.2. In Vivo Methods

In general, in vitro data cannot reflect the variable parameters of healthy and diseased tissues. In vivo methods are therefore required to obtain more accurate measurements. However, these methods are bounded by a number of safety and practicality constraints. For instance, the boundary conditions of the tissue are complicated and tissue volume and position disturbances occur due to cardiac or pulmonary motions [26]. Furthermore, it is required that the acquisition devices do not cause any tissue damage or trauma, disrupt or interfere with the surgical procedure being performed. They should also be usable within restricted environments such as those found in MIS procedures. In general, in vivo tissue property measurement can be divided into invasive and non-invasive methods.

Invasive techniques include indenting, extending or manipulating the tissue or organ surface. In indentation approaches, the stress/strain relationship is determined through placing compressive force on tissue surface by using an indenting device and measuring the depth of indenter penetration. In tissue aspiration techniques, a tube is placed in contact with the tissue surface then a weak vacuum is applied and carefully increased while the deformation of the surface tracked. By assuming an axisymmetric homogeneous tissue area, it is sufficient to measure the profile of the deformed tissue by using a small inclined mirror beside the aspiration hole and a camera. Tissue aspiration enables well-defined boundary conditions to be defined, thus allowing accurate mechanical models to be fitted to the acquired data. Other in vivo measurement techniques use force and position sensors installed on surgical instruments to determine tissue response to manipulations such as grasping or clamping. Force sensors can also be placed in surgical gloves. By connecting to the recording apparatus, these sensors link tissue-instrument interactions with the generated forces.

Non-invasive techniques are generally known as elastography, i.e. elasticity imaging, and are based on applying known displacements or vibrations to the exterior of the tissue and imaging the strain field within. The strain field, which is mainly a function of tissue displacement and material elasticity, is measured before and after, or during the process by using a non-invasive imaging modality such as MRI or ultrasound [27].

### 5.3.3. Validating Measured Parameters

It is essential to validate measured tissue properties and deformable models. Typically, Finite Element Methods (FEM) are used for such verification by observing the deformation behaviour when force is applied in real and simulated experiments. However, it is generally difficult to evaluate deformation accuracy since several factors such as the accuracy of the deformable model, measured tissue parameters, and the acquired geometrical model all come into play. A method for validating deformations with the aim of providing 'gold standard' deformation was proposed by Kerdok et al. [28]. In this experiment, a 3D imaging modality such as CT is used to acquire scans of material samples under normal and different loading states. Image processing is used to automatically track the displacements of spherical fiducials that are uniformly embedded within the sample. The displacements can then be compared against those obtained from computer simulation approaches to allow for more truthful comparison especially for large deformations. For some surgical procedures, soft tissue parameter can be evaluated by using registration techniques [29]. In this case, the tissue surface is scanned pre- and postoperatively and the outcomes of the simulated and real procedures are compared to judge the precision of the parameters derived. Registration is required in order to accurately align the pre- and postoperative scans.

### 5.4. Haptic Rendering

Haptic rendering refers to the process of conveying forces generated in the VE to the user through a haptic interface device. It allows for the manipulation of virtual objects and the perception of their physical characteristics such as shape, mass, deformability, and surface roughness. Surgeons use these essential cues in order to perform surgical procedures. For example, haptic sensation enables surgeons to carry out precise dissections and exposures while avoiding harming surrounding structures [30]. Haptic feedback includes tactile feedback, i.e. touch sensed by high bandwidth receptors, and force feedback, i.e. force sensed by deep low bandwidth receptors. Both components should be used in the simulator to offer high fidelity sensory feedback.

In order to compute haptic forces, information about the objects in the VE and the avatar, which is a representation of the haptic interface device within the VE, is required. Such information includes object position, velocity, acceleration, stiffness and surface texture. In general, a haptic rendering algorithm can be decomposed into three principal components: collision detection, force response, and control algorithms [31]. The first component is used to determine the interaction forces between the avatar and the virtual organs where the computational complexity is determined by the geometry of the avatar. In practice, however, the types of reaction forces perceived by the trainee are constrained by the capabilities of the haptic interface device. For instance, glove-based devices can only provide tactile force feedback while wearable exoskeleton devices can convey more complex forces with multiple Degrees-of-Freedom (DOFs). When a collision is detected, force-response algorithms approximate the generated contact forces by considering the positions of the avatar and all other virtual objects. The control algorithms, on the other hand, work by minimising the difference between the computed response forces and those achievable by the force feedback device, resulting in the actual generated forces. The actual generated forces should also be used by the simulation engine in order to update the information of the virtual objects.

One of the major requirements for realistic haptic rendering is the high feedback update rate. Compared to a relatively low frequency of 30 Hz required in graphics rendering for smooth visual feedback, contact force interactions require frequencies of 1kHz or higher. Otherwise, mushy interactions and system instabilities could occur [32]. In order to meet such constraints, the complexity of computing intersections between simulation models is alleviated by representing the avatar as a point or a collection of points termed the end effectors. In this case, the components of the interaction force have to be computed only at the end effectors, which accelerates collision and force calculations. Such reduction is accentuated by the physical limitations of force feedback devices in terms of DOFs, resolution, and bandwidth.

Additional techniques can also be used to maintain high haptic feedback refresh rate. For example, a well-known bottleneck in surgical simulation occurs when haptic rendering is coupled with computationally intensive tasks such as checking for collisions. Consequently, the update frequency of the physical objects being simulated typically ranges from 20Hz to 150Hz [33]. The difference between computed and required haptic feedback frequencies can be equalised by using extrapolation, operating on models with reduced complexity, and using localised model representations. The stability of the haptic system can be further increased by decoupling haptic device control from the generation of the simulation environment.

Although a number of techniques have been proposed for modelling haptic interactions with implicit surfaces and NURBS representations, the majority of haptic rendering algorithms consider objects with simple planar polygonal models where force interactions can be efficiently resolved. In this case, the force normal to the plane is computed proportional to material stiffness, whereas the force tangential to the plane is proportional to velocity (material damping). Moreover, in order to prevent the end effectors from penetrating through planar models, a problem that is due to the limited mechanical stiffness of haptic devices, several techniques have been proposed. In this case, a virtual location that corresponds to where the haptic interface point would be if the model was infinitely stiff is defined. Then, by having a spring connecting the virtual location and the end effectors, realistic force interactions can be computed without the effectors penetrating the model.

Further issues should be considered when planar polygons are used for haptic calculations. For example, with planar polygons, curved objects will feel faceted, and for concave objects, there is an ambiguity in determining the face to be used for collision calculation. Furthermore, the objects will lack haptic realism in terms of surface texture and friction. The last two properties are of special importance in surgical simulation where the tactile sensation through palpating, stroking, or indenting the organ surface, is used extensively by the surgeon to judge tissue conditions. In existing research, several force shading algorithms have been proposed to decrease the faceted effect resulting from force discontinuities along polygon edges. This, in a way, is analogous to the Phong shading algorithm used in computer graphics. A number of haptic texture approaches have also been presented for local surface-dependent force rendering [34, 35]. However, due to the lack of appropriate hardware, tactile feedback has not yet been widely used in surgical simulation. Moreover, the complex interaction found in surgical procedures is difficult to model with the current approach of using point effectors [36].

## 5.5. Image Synthesis

Visual perception plays a major role in the success of surgical procedures. It is the primary information channel available to surgeons who depend on scene appearance to determine tissue properties, instrument pathways and necessary interactions. Therefore, photorealistic rendering is an important aspect of the simulation environment. Realistic visualisation in the simulation environment is achieved by using computer graphics techniques designed for visual realism. This involves a number of topics such as object representation, modelling, illumination, surface properties, shadows, anti-aliasing and colour perception. Since all organ surfaces are covered with micro-structures that provide information about tissue characteristics, texturing provides an essential visual cue in addition to its use for increasing realism. As pointed out by Szekley et al. [15], 3D space orientation and depth cues can be given by texture perspective shrinkage, and its visual appearance can indicate tissue pathology. The authors have proposed a technique for texture generation that overcomes the fundamental problem of using limited patient-specific texture for training by providing a large synthetic texture database for organs with different pathologies. In this case, a combination of procedural textures is used to account for texture variations at organ surface.

In addition to texturing, specular highlights constitute another important visual cue in endoscopic procedures. They provide shape curvature, distance, as well as surface contact cues. In conventional graphics rendering, specular highlights are typically computed by using the Phong reflection model [37] where the lighting equation is evaluated at object vertices. However, specular highlights are generally smaller than the facets of the geometric model and this can result in specular aliasing. Neyret et al. [38] presented a method to alleviate these problems and to provide realistic specular reflection in surgical simulation by using an environment texture to represent the specular spots. The method takes into account light intensity variations due to distance changes and the effects of surface roughness on the shape of the generated highlights.

Image Based Rendering (IBR) approaches have also been proposed for modelling and rendering deformable soft tissues with a high degree of photorealism. For instance, cylindrical relief texture mapping [39] had been used to increase realism of textured cylindrical surfaces by adding 3D details and enhancing the depth conceived by the viewer by introducing motion parallax. In [4], a method for simulating soft tissue deformation with IBR is described based on the association of a depth map with the texture image and the incorporation of micro-surface details to generate photorealistic images representing soft tissue deformations. Micro-surface details are augmented to the model with 3D image warping to drastically reduce the polygonal count required to model the scene whilst preserving deformed small surface details to offer a high level of photorealism.

In minimally invasive surgery, specular reflections provide an important visual cue for tissue deformation, depth and orientation. [5] and [40] describe a photo-realistic rendering approach based on real-time per-pixel effects by using the graphics hardware where noise functions can be used to control the shape of the specular highlights. Improved realism is achieved by a combined use of specular reflectance and refractance maps to model the effect of surface details and mucous layer on the overall visual appearance of the tissue.

## 6. Classification of Simulators

The diversity of the surgical procedures along with the rapid changes in technology means that no single simulation environment can cater for the general needs of surgical simulation. While some simulation systems are designed to teach surgeons how to perform straightforward tasks such as needle placement and basic hand-eye coordination skills, others offer advanced training on more complex procedures. Simulators also vary by the degree of visual realism, the availability of automatic scoring and assessment, and the computational requirements dictated by modelling and simulation complexity. In general, surgical simulators can be classified according to the complexity of the procedures being practiced as [41]:

- Simple single-task simulators: these simulators focus on teaching basic surgical tasks with varying levels of difficulty such as instrument manipulation, needle or catheter placement, and endoscope navigation. They are characterised by a simple computational model that allows for realistic visual and haptic feedback. Some simulators can train different instrument manoeuvres and case scenarios. They also permit the use of additional supporting devices such as biopsy forceps. Examples in this category include intravenous, colonoscopy, and bronchoscopy simulators.

- Complex single-task simulators: some of the surgical tasks are complex with regard to the number of required instrument manipulations. For instance, suturing and dissection are two tasks requiring advanced computational models and complicated instrument movements. They may involve more than one device interacting with each other and acting simultaneously on the same tissue. Sophisticated modelling and device interfaces are needed in order to account for realistic haptic interactions. Accurate tissue deformation parameters and texture information are essential to create high quality visual realism. In order to meet the demand for simulating complex tasks, the use of specialised hardware and parallel architecture have been exploited. This is reflected in the high cost required to deploy and maintain the technology. In addition, the simulators have to trade certain aspects, such as visual realism, for realistic haptic feedback. Examples in this category include anastomosis and limb trauma simulators.

- Multiple-task simulators: these simulators combine more than one task to train a complete process. If these tasks are complex, then the immense simulation requirements render the whole modality computationally impractical unless crude models are used. This can result in low-fidelity training for each of the individual components [42]. Therefore, with the current technology, it is more efficient to identify and train critical tasks of complex procedures independently. However, multi-task simulators have the advantage of decreased cost and the possibility of being used by more than one surgical group.

## 7. Skills Training and Assessment

### 7.1. Surgical Skill Components

It has long been recognised that computer-based surgical simulation provides an important means of training as well as assessing skills. A number of methods for identifying surgical skills are based on analysing the steps of the task being performed. For example, in task and motion analysis [43], a number of laparoscopic tasks such as suturing, knot tying, cutting and dissection were considered. Videotape timeline analysis was used to decompose each task into various subtasks, followed by defining a number of motion and force primitives for each subtask. It has been shown from these studies that hand-eye coordination is an important component for laparoscopic surgery and that time can be used as a metric of performance index. Virtual training scenarios can then be developed to specifically train certain recognised skills. Other methods for identifying the components of surgical skills include consultation with content-matter experts and error analysis. Overall, the range of fundamental skills that can be taught by using virtual environments include [44]:

- Perceptual-motor skills: surgeons depend on visual and tactile cues such as texture appearance and contact differences to carry out open procedures. In MIS, these cues are greatly diminished and surgeons adjust to this inherent limitation by developing special perceptual motor skills and psychophysical adaptation such as conscious-inhibition. Simulation environments can be used to train as well as develop these skills.

- Spatial skills: although important, some surgical skills that depend on spatial cognitive ability are hard to define. For example, to obtain proper exposure, the surgeon adjusts the tissue for access and relies on his/her ability to classify objects according to their spatial characteristics such as location, movement, extent, shape and connectivity. MIS procedures such as laparoscopy depend extensively on spatial ability since less perceptual information is available. It has also been shown that the visual-spatial ability is directly proportional to the efficiency of hand motion for successful surgical performance.

- Procedural skills: these are the skills acquired through the regular practice of a certain surgical procedure. Simulation can be used to enhance the procedural skills by emphasising the critical steps of a procedure and introducing different complexities that trainees have to handle.

It is worth noting that different degrees of visual realism are required for training different surgical skills, and that the practice of a set of abstract tasks can improve certain skills. For example, a computer-based skills assessment device can be used to train object acquisition, traversal and manipulation. It can also differentiate the different skill levels of experienced and novice laparoscopic surgeons. More complex tasks such as suturing and dissection can also be taught with simplistic environments [45]. However, for training skills with a higher cognitive load, such as spatial reasoning, a key concern is to have enough level of realism, so the simulated environment is as close as possible to the real one. This is required in order to facilitate the transfer of skills between the artificial and real situations.

### 7.2. Metrics for Skill Assessment

For the assessment of surgical skills, an ideal procedure should include a number of features such that it must be: (a) valid and reflect true measurements, (b) objective and reproducible, and (c) fair and direct [7]. In general, assessment techniques can be divided into subjective and objective methods. Subjective or qualitative assessment is the commonly used approach built on the well-established apprenticeship surgical training model. In this case, an expert surgeon observes the pattern of hand and instrument movements and evaluates the outcome of a surgical procedure. Therefore, based on personal opinion, the capabilities of the trainees are qualitatively judged. A number of objective skills assessment approaches have been proposed [46] where pre-trained observers judge trainees performing specific structured procedures in laboratory settings. A common problem associated with this type of assessment is that the monitoring and judgement process is time-consuming and labour intensive [7].

One of the potential advantages of surgical simulation is to assess the skills of surgeons objectively or quantitatively. Instrument movements are tracked and analysed in order to establish relationship with the level of expertise. The skill level of trainees is enumerated by comparing their performance and dexterity in the simulator environment with those of experts. Specifying the parameters needed for an assessment model should consider the method by which expert surgeons evaluate trainees, which is usually dependent on the task being performed. However, a number of task-independent components for laparoscopic surgical competence have been identified. Factors such as the compactness of the spatial distribution of instrument tip movement, total number of movements, motion smoothness, depth perception, and time for task completion are typically considered in order to differentiate between trainees. Five different metrics based on kinematic analysis, as shown in Table 1, are used to quantitatively measure these components [47].

In practice, skill assessment techniques should focus not only on evaluating performance, but also on the proficiency by which the procedures are carried out. It is also important to take into account the clinical consequences of decisions made during simulation. Some methods utilise simple force information, such as maximum, minimum, and integrals along with collision computation to evaluate the pressure imposed on the tissue during training sessions [48]. Therefore, in addition to the previous metrics, a force-based quality measure can also be defined for evaluating surgical skills. However, defining, training and assessing higher cognitive skills are difficult to conduct [36]. In general, assessing surgical competency is highly complex and the basic precepts of surgical training assessment, i.e. cognitive and psychomotor, should all be considered in order to determine the associated skill level [49].

**Table 1** Five different metrics based on kinematic analysis are used to quantitatively measure the components of surgical skill level.

| Metric | Description |
| --- | --- |
| Time | Time needed to carry out the surgical task |
| Path length | Distances travelled by instruments and their spatial relations |
| Motion Smoothness | Changes in instrument acceleration |
| Depth Perception | How deep the instruments travels |
| Response Orientation | How much instrument rotation is needed before it is properly aligned |

## 8. Challenges to Surgical Simulation

The advantages of using simulation environments for teaching surgical skills have been reported in several studies [50, 51, 52]. Moreover, its potential use for assessment and certification is becoming well recognized by a large number of professional medical societies. However, as pointed out by [49], for successful surgical simulation, a number of challenges, such as enhancing technical fidelity and standardising assessment metrics, should be taken into account. Accurate deformation models, which incorporate precise material properties, are also needed whereas real-time interactions should be enabled with a high degree of visual and haptic realism. Moreover, improved haptic devices that allow for more DOFs and different types of interaction, e.g. provide for tactile as well as force feedbacks, are required. Finally, assessment methods and the metrics used should be standardised since there are still no uniform tests or reporting schemes available, which makes it difficult to determine the correspondence between different training approaches. Additional challenges are related to introducing rigorous evaluation of simulation reliability and validity, and the integration of surgical simulation into medical educational curriculum.

### 8.1. Reliability and Validity

Although the development of surgical simulation has been the focus of research for many years, a relatively small number of studies are available for assessing the effectiveness of this modality. Most of the available work compares performance of subject groups with different skill levels in the simulator environment [53] and several experiments have shown an improvement in hand-eye coordination as a result of practicing on simulators. However, the nature of skill transfer from simulated to real environments is task dependent and still the subject of ongoing research. Moreover, most of the reported results are from experiments performed in restricted environments. In general, for a new measurement tool such as surgical simulation, clear evidence for two important issues, i.e., reliability and validity, has to be provided before the results generated can be interpreted with confidence [54].

Reliability denotes the concept of the precision of measured parameters. Since several approximations are considered in the simulation in terms of model representation, deformation behaviour, and visual and haptic rendering, it is necessary to introduce methods to verify the accuracy of the overall simulation.

Training and assessment validity is another important topic in surgical simulation. The relationship between the procedures trained or evaluated within the simulation environment and those carried out in reality has to be established. Ideally, before a simulator is used to train or assess specific skills, it has to go through a validation phase where a number of validity measures are examined [36, 54]:

- Content validity: the skills trained or assessed in the simulation have to be related to the core skills of the task being practiced. The scoring system should record and measure trainees' performance and proficiency. In addition, the simulator contents must be correct in terms of appearance and spatial relationships to avoid providing trainees with misleading cues.

- Criterion validity: the performance in the simulation and real environment should be correlated. This also has to do with the implicit surgical experience and the skills transferred from synthetic to real environments.

- Construct validity: for construct validity to be achieved, experienced surgeons must perform better in the simulation environment, which means the simulation has the capacity to measure and differentiate between different skill levels.

## 8.2. Incorporation within Medical Curriculum

Surgical simulation is becoming a widely accepted modality for teaching surgical skills and many simulators are now commercially available. However, surgical simulation still has not been integrated into surgeons' education and training curriculum. The hurdles to accepting VE-based surgical training by the surgical community can be summarised as following [55, 56]:

- Lack of realism: even though it is shown that abstract training can lead to certain skill improvements, existing simulators still cannot provide a convincing immersion due to technology limitations.

- Lack of appropriate assessment methodologies: current assessment metrics used in simulators lack clinical significance. They measure some performance characteristics and don't consider their relation to required real world skills. Moreover, the outcome of surgical procedure in terms of success or failure should also be considered in the assessment.

- Lack of appropriate contents: developing the contents required for using surgical simulation as an educational tool is necessary to enable the use of this new modality.

Of all obstacles, developing the content and curriculum for the simulation is the most difficult challenge [49]. In order for surgical simulation to be successfully integrated within the medical curriculum, appropriate contents that match the capabilities of exiting technology should be devised. Existing technologies, such as simple-task simulators and real-time visually realistic navigation systems, can be incorporated into current curriculum to train simple dexterity tasks and teach basic anatomy and vessel structures. As the technology advances, intricate tissue interactions and complex skills training can be gradually adopted.

Having said that, recent studies have shown that medical students/surgeons using simulators perform better and retain more of what they learned than their colleagues who use more traditional methods of medical training. Evidence-based laparoscopic simulation curriculum shortens the clinical learning curve and reduces surgical adverse event [57]: This study compares the effectiveness of a proficiency-based preclinical simulation training in laparoscopy with conventional surgical training and conventional surgical training interspersed with standard simulation sessions. It concluded that proficiency-based preclinical training has a positive impact on the learning curve of a laparoscopic cholecystectomy and diminishes adverse events.

Laparoscopic simulation trainers offer pedagogical advantages over other medical education formats [58]: Because they are learner centric, laparoscopic simulators allow for skill acquisition and improvement without the responsibility of patient care. Ongoing research is providing new evidence that simulation is of paramount importance to continuing professional development as it has been in graduate medical education.

The role of simulation in the development of endovascular surgical skills [59]: As the scope for endovascular therapy increases, due to the rapid innovation, evolution and refinement of technology, so too do patients' therapeutic options. This climate has also opened the door for more novel training adjuncts, to address the gaps that exist in our current endovascular training curriculum. New advances in technology mean simulators can continue to provide an important training adjunct for even the most experienced practitioners. However, benefit gained is dependent on the choice of simulation model. It was found out that virtual reality simulation is the most promising, but cost remains a prohibitive factor.

The emerging role of screen-based simulators in the training and assessment of colonoscopists [60]: High fidelity screen-based simulators hold great appeal for endoscopy training. Potentially, their incorporation into endoscopy training curricula could enhance speed of acquisition of skills and improve patient comfort and safety during the initial phase of learning. They could also be used to demonstrate competence as part of the future relicensing and revalidation of trained endoscopists.

## 9. Technology Providers

From trainee surgeons to established surgeons learning a new technique, a surgery simulator can be an excellent way to learn in a low-risk environment. Synthetic/Mannequins and VE-based simulators can be combined to provide both visual and haptic feedback resulting into enhanced immersion/realism and better surgical simulation experience. We next describe key VE-based simulation providers (in random order).

**Medical Realities** [61]: Medical Realities is a Technology Enhanced Healthcare Company. Enabling the Convergence of immersive Technology and Validated content to Improve Healthcare. The Medical Realities Platform delivers high-quality surgical training using Virtual Reality. Trainees are immersed while watching an experienced surgeon perform a surgical procedure. Depending on the procedure, this can include laparoscopic or microscope feeds, and a 3D close-up feed of the area that is being operated on.

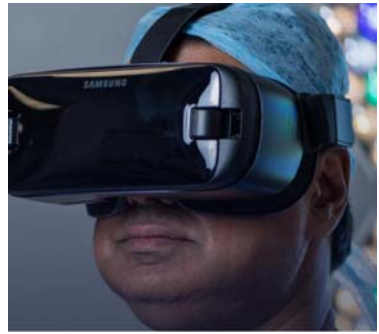

VR Anatomy platform lets trainees explore in detail, the anatomy required for each procedure where they can toggle back and forth between the 360° video and the anatomy at any point.

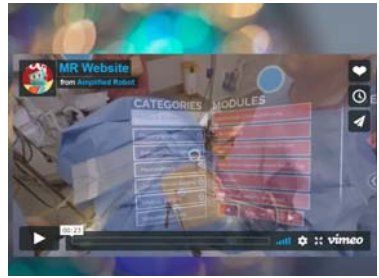

A question bank for each module which is written by teaching consultants is used for trainee evaluation.

**Immersivetouch** [62]: The company offers VR Simulator for Surgeons. Two platforms are offered: the ImmersiveView Surgical Planning and ImmersiveTraining. ImmersiveView Surgical Planning (IVSP), is a platform that generates high-fidelity 3D VR replicas from patient DICOM scan data and provides a variety of tools for use directly on the model in VR. This allows surgeons to study, assess, and plan their surgeries, collaborate intra-operatively with their surgical team, better educate patients about their upcoming surgery.

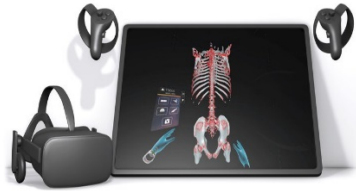

**EOSurgical** [63]: Provides evidence-based, accessible simulator that tracks performance and training.

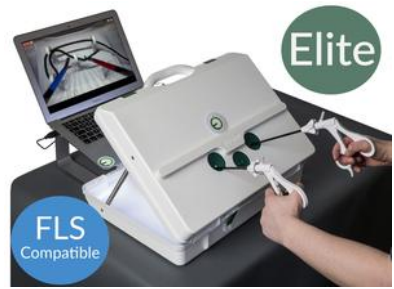

**zSpace** [64]: zSpace provides an immersive learning platform for medical training, allowing students to examine virtual living body structures with accurate anatomical representation, plan procedures, and present findings. Their solution features Human Anatomy Atlas which is a human anatomy general reference used to explore human body systems, anatomical structures, musculoskeletal animations, and quiz questions. Vizitech ECG allows students to practice ECG electrode placement, understand relationships between electrode placement and the ECG strip, and study abnormal ECGs.

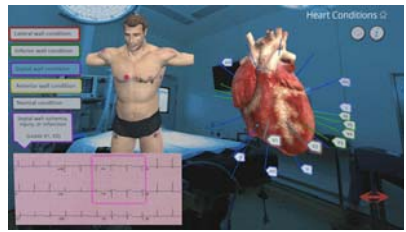

**OramaVR** [65]: Offers VR-based surgical simulation solutions for Total Knee and Hip Arthroplasty operations, Multi-Surgeon VR, and VR psychomotor software development kit. Episodes include OR, Incision, Drilling and Implants.

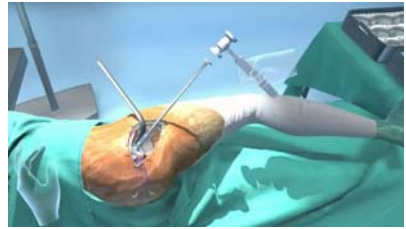

**3D Systems** (formerly Simbionx) [66]: the company offers a wide range of simulation, training and education solutions for medical professionals and the healthcare industry. Their products train surgeons to perform Minimally Invasive Surgery (MIS) and interventional procedures. They cover procedures and examinations in arthroscopic, endovascular, pelvic, spine and bronchoscopy, among others.

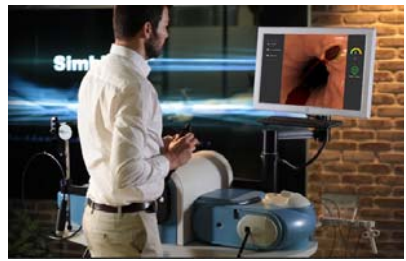

## 10. Discussion and Conclusions

In this article we discussed different surgical training approaches that are currently available. We have explained the limitations of conventional training and how the emergence of new surgical procedures and increasing surgical complexity have led to new training and assessment techniques. The use of surgical simulation in surgical training and evaluation has a number of advantages. For instance, it can be used to enhance surgical skills by allowing repeated practice and to maintain acquired level of competence. Furthermore, it is a more efficient and cost-effective modality that poses no risk to patients and avoids many ethical and legal complications.

For effective training and assessment, several technical issues, mostly related to realism and accuracy, have to be considered in surgical simulation. Since surgeons mainly depend on perceptual-motor coordination and contact to carry out procedures, the simulator must provide essential visual as well as haptic cues that closely resemble those found in real situations. Precise anatomy and tissue deformation are required if the system is to be used for planning or determining the outcome of a surgical procedure. Therefore, technical challenges include the development of accurate tissue property acquisition and deformation models that can accommodate large non-linear deformations as found in real procedures. More complex behaviours such as bleeding and contractions should also be included in the simulator.

The significance of surgical simulation for skills training and assessment is widely recognised by the medical community. Several skills such as perceptual-motor, spatial and procedural skills can be trained by using VE-based simulators. Moreover, active research is currently being carried out to ensure that the skills taught by the simulator are transferable to the real clinical settings. Surgical simulation also offers an effective method for quantitative skill evaluation, thus alleviating the problem of subjective assessment and simplifying the establishment of standards for surgical competency. For this purpose, a number of assessment metrics based on time for task completion and economy of instrument movements have been proposed.

Surgical simulation can facilitate the development of novel surgical techniques by practicing and evaluating procedures in the simulation environment. It is also endorsing other trends in surgery. For instance, simulation environments can be used to improve telemedicine or telepresence applications that deliver medical care to distant locations. However, the main limiting factors for the large-scale adaptation of surgical simulation in hospitals and teaching institutions are primarily due to the high cost of developing and maintaining this high-tech solution. In fact, as pointed out by Satava [49], the financial issue may cause the greatest delay for simulation acceptance. This results from the lack of appropriate business model that motivates companies to develop surgical simulators along with the relatively small market size. Furthermore, to be financially successful, the simulator must be capable of offering the following features: (a) multifunctional training for many different specialities, (b) have the ability to train different levels of expertise, and (c) must be integrated into several aspects of the medical practice such as preoperative planning, research and virtual prototyping of instruments or equipment. Although these requirements seem ambitious, it is anticipated that the cost will go down as the technology advances and more economically feasible solutions will become available.

In conclusion, based on the overwhelming evidence for the usefulness of simulation-based training in clinical skills education, it is advisable that simulation must be integrated in the medical training curriculum as distributed training sessions with the possibility of directed, self-regulated learning in professional training facilities. Simulation-based training to proficiency should be mandatory before trainees are allowed to perform procedures on patients [67]. Furthermore, simulation programme leadership must set objectives, plan and pilot content, budget equipment and personnel, evaluate effectiveness, and secure institutional support in order to sustain a successful simulation-based medical education programmes [68].